\documentclass[%
 reprint,
 amsmath,amssymb,twocolumn,
 aps,pra,superscriptaddress,tightenlines
]{revtex4-2}

\usepackage{pifont}
\usepackage{graphicx}
\usepackage{dcolumn}
\usepackage{bm}

\usepackage{marvosym}
\usepackage{braket} 
\usepackage[colorlinks=true,
            linkcolor=black, 
            anchorcolor=black,
            citecolor=black, 
            urlcolor=blue
            ]{hyperref}

\usepackage{color}
\usepackage{mathrsfs}
\usepackage{slashed}
\usepackage{appendix}

\begin{document}

\preprint{APS/123-QED}

\title{Experimental  violation of Leggett-Garg inequality in a three-level trapped-ion system}

\author{Tianxiang Zhan}
\thanks{These authors contributed equally to this work.}%
\affiliation{%
Institute for Quantum Science and Technology, College of Science, National University of Defense Technology, Changsha Hunan 410073, China
}%
\affiliation{%
Hunan Key Laboratory of Mechanism and Technology of Quantum Information, Changsha Hunan 410073, China
}
\author{Chunwang Wu}
\thanks{These authors contributed equally to this work.}%
\affiliation{%
Institute for Quantum Science and Technology, College of Science, National University of Defense Technology, Changsha Hunan 410073, China
}%
\affiliation{%
Hunan Key Laboratory of Mechanism and Technology of Quantum Information, Changsha Hunan 410073, China
}
\author{Manchao Zhang}
\affiliation{%
Institute for Quantum Science and Technology, College of Science, National University of Defense Technology, Changsha Hunan 410073, China
}%
\affiliation{%
Hunan Key Laboratory of Mechanism and Technology of Quantum Information, Changsha Hunan 410073, China
}
\author{Qingqing Qin}
\affiliation{%
Institute for Quantum Science and Technology, College of Science, National University of Defense Technology, Changsha Hunan 410073, China
}%
\affiliation{%
Hunan Key Laboratory of Mechanism and Technology of Quantum Information, Changsha Hunan 410073, China
}
\author{Xueying Yang}
\affiliation{%
Institute for Quantum Science and Technology, College of Science, National University of Defense Technology, Changsha Hunan 410073, China
}%
\affiliation{%
Hunan Key Laboratory of Mechanism and Technology of Quantum Information, Changsha Hunan 410073, China
}
\author{Han Hu}
\affiliation{%
Institute for Quantum Science and Technology, College of Science, National University of Defense Technology, Changsha Hunan 410073, China
}%
\affiliation{%
Hunan Key Laboratory of Mechanism and Technology of Quantum Information, Changsha Hunan 410073, China
}
\author{Wenbo Su}
\affiliation{%
Institute for Quantum Science and Technology, College of Science, National University of Defense Technology, Changsha Hunan 410073, China
}%
\affiliation{%
Hunan Key Laboratory of Mechanism and Technology of Quantum Information, Changsha Hunan 410073, China
}
\author{Jie Zhang}
\affiliation{%
Institute for Quantum Science and Technology, College of Science, National University of Defense Technology, Changsha Hunan 410073, China
}%
\affiliation{%
Hunan Key Laboratory of Mechanism and Technology of Quantum Information, Changsha Hunan 410073, China
}
\author{Ting Chen}
\affiliation{%
Institute for Quantum Science and Technology, College of Science, National University of Defense Technology, Changsha Hunan 410073, China
}%
\affiliation{%
Hunan Key Laboratory of Mechanism and Technology of Quantum Information, Changsha Hunan 410073, China
}
\author{Yi Xie}
\affiliation{%
Institute for Quantum Science and Technology, College of Science, National University of Defense Technology, Changsha Hunan 410073, China
}%
\affiliation{%
Hunan Key Laboratory of Mechanism and Technology of Quantum Information, Changsha Hunan 410073, China
}
\author{Wei Wu}%
 \altaffiliation{weiwu@nudt.edu.cn}
\affiliation{%
Institute for Quantum Science and Technology, College of Science, National University of Defense Technology, Changsha Hunan 410073, China
}%
\affiliation{%
Hunan Key Laboratory of Mechanism and Technology of Quantum Information, Changsha Hunan 410073, China
}
\author{Pingxing Chen}
\affiliation{%
Institute for Quantum Science and Technology, College of Science, National University of Defense Technology, Changsha Hunan 410073, China
}%
\affiliation{%
Hunan Key Laboratory of Mechanism and Technology of Quantum Information, Changsha Hunan 410073, China
}

\date{\today}

\begin{abstract}
Leggett-Garg inequality (LGI) studies the temporal correlation in the evolution of physical systems. Classical systems obey the LGI but quantum systems may violate it. The extent of the violation depends on the dimension of the quantum system and the state update rule. In this work, we experimentally test the LGI in a three-level trapped-ion system under the model of a large spin precessing in a magnetic field. The Von Neumann and Lüders state update rules are employed in our system for direct comparative analysis. The maximum observed value of Leggett-Garg correlator under the Von Neumann state update rule is $K_3 = 1.739 \pm 0.014$, which demonstrates a violation of the Lüders bound by $17$ standard deviations and is by far the most significant violation in natural three-level systems.
\end{abstract}

\maketitle


\section{\label{sec:level1}Introduction}

One of the most counterintuitive features of quantum mechanics is the superposition of quantum states \cite{glauber1963quantum}. In 1935, Schrödinger extended the superposition from the micro-world to the macro-world through the thought experiment of Schrödinger's cat \cite{schrodinger1935gegenwartige}. However, it seems to contradict the fact that objects in our daily life are always in a certain state. According to this contradiction, in 1985, Leggett and Garg proposed the Leggett-Garg inequality (LGI) based on two fundamental assumptions of the macro-world \cite{leggett1985quantum}: (A1) Macroscopic realism (MR): A macroscopic system is always in one of the macroscopically distinct states; (A2) Noninvasive measurability (NIM) at the macroscopic level: A determination method can be found that does not affect the past and the future of the system \cite{leggett1985quantum}. Because these two assumptions are invalid in quantum mechanics, quantum systems may violate the LGI.

Based on above two assumptions, the standard LGI is
\begin{eqnarray}
-3\leq K_3\leq 1, \label{con:K3}
\end{eqnarray}
where $K_3$ is a linear combination of temporal correlations of observables measured sequentially at different moments \cite{budroni2014temporal}. The upper bound 1 is usually called the classical bound (CB), which limits the behavior of classical systems. For two-dimensional quantum systems, the maximum value of $K_3$ is 1.5 \cite{leggett1985quantum}, which is usually called the Lüders bound (LB) \cite{budroni2014temporal} (this bound is also called the temporal Tsirelson bound (TTB) \cite{cirel1980quantum,poh2015approaching,navascues2008convergent}). For higher dimensional quantum systems, it has been strictly proved that the maximum value of $K_3$ is still 1.5 \cite{budroni2013bounding} under the Lüders state update rule (LSUR) \cite{luders2006concerning}. Later, Budroni and Emary proved that $K_3$ might exceed 1.5 in systems with three or more dimensions under the Von Neumann state update rule (VSUR) \cite{von2018mathematical}, and the maximum value of $K_3$ increases with  increasing dimension, which can approach 3, the algebraic bound \cite{budroni2014temporal}.

In 2010, Palacios-Laloy announced the first experimental violation of LGI \cite{palacios2010experimental}. Subsequently, several experimental tests of LGI in two-level quantum systems were reported, including the single photon system \cite{xu2011experimental}, the spin-bearing phosphorus impurities in silicon system \cite{knee2012violation}, the nuclear magnetic resonance system \cite{katiyar2013violation}, the superconducting system \cite{santini2022experimental}, etc. The first violation of LGI in a three-level system was reported by George et al. They implemented their experiment under the LSUR in a nitrogen-vacancy center in diamond system \cite{george2013opening}. In 2017, two experiments that violate the LGI were realized in the nuclear magnetic resonance system \cite{katiyar2017experimental} and the single photon system \cite{wang2017enhanced} under the VSUR. They both used two qubits to mimic a qutrit. In 2022, Maimaitiyiming et al.\ used a natural three-level system to test the LGI in a nitrogen-vacancy center in diamond system \cite{tusun2022experimental}. However, they only collected one data point, and the experimental value is not very consistent with the ideal theoretical value subject to the short coherence time of this system. Moreover, all experiments in three-level systems mentioned above only adopted one state update rule and did not give a direct comparison of $K_3$ under two state update rules experimentally.

In this work, we experimentally test the LGI in a natural three-level trapped-ion system under the evolution model of a large spin precessing in a magnetic field \cite{budroni2014temporal}. Under this model, We obtained the most significant violation in a natural three-level system benefiting from the high-fidelity operations and long coherence time of the trapped-ion system.  Moreover, for the first time, the Von Neumann and Lüders state update rules are employed in LGI research for direct experimental comparative analysis. Our results show different upper bounds of LGI under two state update rules.

\section{theoretical model}
\begin{figure}[h]
\includegraphics[width=0.43\textwidth]{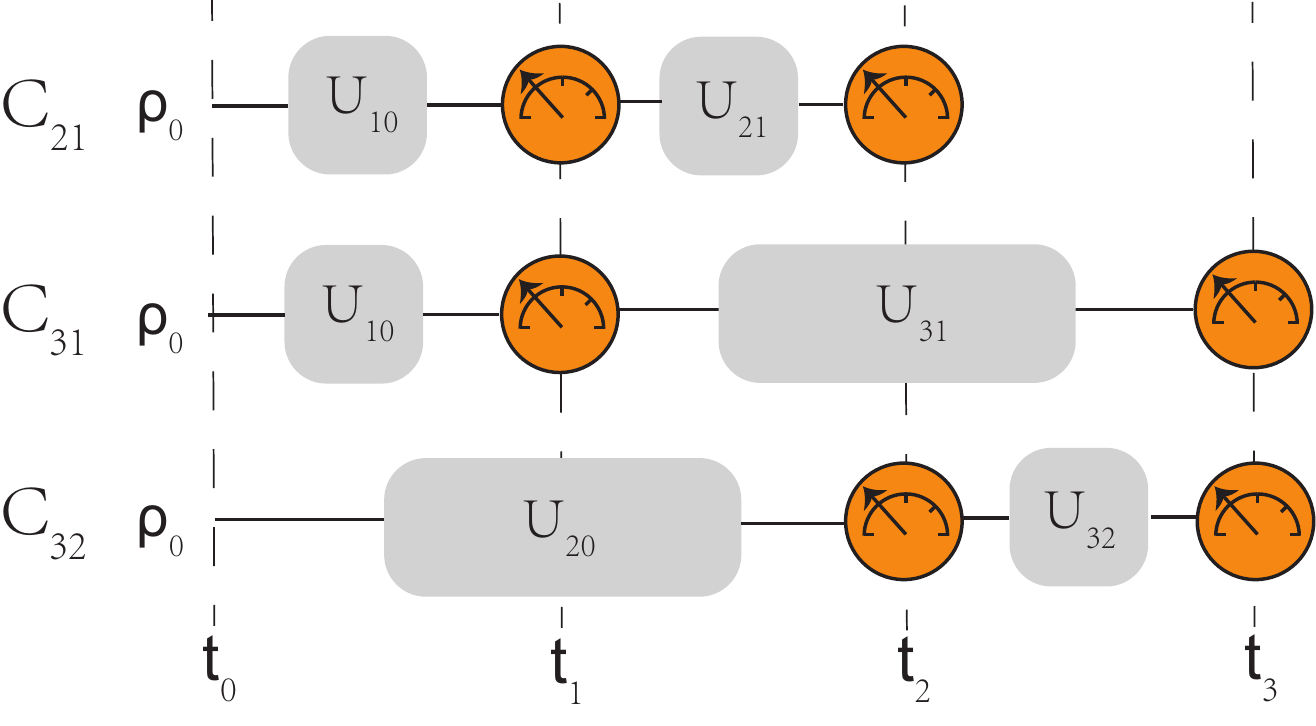}
\caption{\label{fig:3}  Scheme for the LGI test. $\rho _ 0$ represents the initial density operator. Three separate experiments need to be performed to obtain $C_{21}$, $C_{32}$ and $C_{31}$. The $Q$ values are measured at the corresponding two moments, and each experiment is performed multiple times to obtain the joint probability $P_{ij}(Q_i,Q_j)$.}
\end{figure}

LGI considers the temporal correlation of system evolution. We can assume that there exists a dichotomous observable quantity $Q=\pm1$ in a macroscopic system. Due to the MR assumption, a system’s state can only take a definite ontic state corresponding to $Q=+1$ or $Q=-1$. The correlation function between moments $t_i$ and $t_j$ is
\begin{eqnarray}
C_{ij}=\sum_{Q_i,Q_j=\pm 1} Q_i\,Q_j\,P_{ij}(Q_i,Q_j),
\end{eqnarray}
where $P_{ij}(Q_i,Q_j)$ represents the joint probability of obtaining the measurement outcomes $Q_i$ and $Q_j$ at moments $t_i$ and $t_j$. Three correlation functions $C_{21}$, $C_{31}$ and $C_{32}$ can be defined by selecting three measurement moments $t_1$, $t_2$ and $t_3$, and then we can define
\begin{eqnarray}
K_3=C_{21}+C_{32}-C_{31}. \label{K3}
\end{eqnarray}
Under the restriction of the NIM assumption, it is easy to derive Eq.\,(\ref{con:K3}) \cite{emary2013leggett}. The measurement of $K_3$ requires three independent experiments, and each experiment selects two of the three measurement moments to obtain $C_{21}$, $C_{32}$, and $C_{31}$. The experimental process is shown in Fig.~\ref{fig:3}.

A quantum system can violate the LGI, and the maximum value of $K_3$ is 1.5 in a two-dimensional quantum system. Regarding systems with three or more dimensions, the maximum value of $K_3$ depends on the state update rule. According to the LSUR, the state after measurement updates following
\begin{eqnarray}
\rho_{L} \mapsto \Pi_{\pm}\rho\Pi_{\pm},
\end{eqnarray}
where $\Pi_+$ and $\Pi_-$ denote projection operators corresponding to $+1$ and $-1$ respectively \cite{budroni2014temporal}. Under the VSUR, the state updates following
\begin{eqnarray}
\rho_{V} \mapsto \sum_k (\Pi^k_\pm\rho\Pi^k_\pm),
\end{eqnarray}
where $\Pi^k_\pm$ denote one-dimensional projectors, and k denotes the degeneracy of eigenvalue $Q$ \cite{budroni2014temporal}. It is easy to see that the measurement under the LSUR only distinguishes eigenspaces corresponding to different eigenvalues but does not distinguish which one-dimensional subspace is projected to. However, the measurement using the VSUR projects each one-dimensional subspace and obtains more information. Therefore, the LSUR respects degeneracy, but the VSUR destroys degeneracy. The maximum value of $K_3$ is 1.5 and is independent of the system's dimension under the LSUR, but it can exceed 1.5 when the VSUR is adopted. Some works claim that this transcendental behavior comes from the fact that the VSUR introduces an additional non-classicality and can not be considered a violation of macrorealism in the usual sense \cite{kumari2018violation}.

In this work, we employ the model of a large spin precessing in a magnetic field  mentioned in Ref.\,\cite{budroni2014temporal} in a three-level system. The corresponding Hamiltonian ($\hbar=1$) can be expressed as
\begin{eqnarray}
H=\Omega J_x,
\end{eqnarray}
where $\Omega$ is the level spacing and $J_x$ is the $x$ component of the angular momentum operator. The correlation function between two moments $t_\alpha$ and $t_{\beta}$ can be written as
\begin{eqnarray}
C_{\beta\alpha}=\sum_{l,m} q_lq_m\text{Tr}\left\{\Pi_mU_{\beta\alpha}\Pi_lU_{\alpha0}\rho_0U^{\dagger}_{\alpha0}\Pi_lU^{\dagger}_{\beta\alpha}\right\},
\end{eqnarray}
where $q_l$ and $q_m$ denote the output results ±1 related to the projection operators $\Pi_l$ and $\Pi_m$ respectively \cite{budroni2014temporal}. The VSUR is adopted when $\Pi_l$ and $\Pi_m$ represent a one-dimensional projection operator, and the LSUR is adopted when $\Pi_l$ and $\Pi_m$ represent the projection operator of the eigenspace corresponding to a certain eigenvalue. $\rho_0$ denotes the density operator at the initial moment $t_0$. $U_{\beta\alpha}$ denotes the unitary evolution operator between the moments $t_{\alpha}$ and $t_{\beta}$ which can be written as
\begin{eqnarray}
U_{\beta \alpha}=e^{-i H\left(t_{\beta}-t_{\alpha}\right)}. \label{Eq8}
\end{eqnarray}
We define three measurement moments $t_1$, $t_2$ and $t_3$, and set the time intervals as
\begin{align}\label{eq2}
\Omega\left(t_{1}-t_{0}\right)=\pi; \\
t_{2}-t_{1}=t_{3}-t_{2}=\tau. \label{Eq10}
\end{align}
According to our experimental settings under the LSUR, $K_3$ reads
\begin{equation}
K_3=-\frac{1}{8}+2 \cos (\Omega \tau)-\cos (2 \Omega \tau)+\frac{1}{8} \cos (4 \Omega \tau) .
\end{equation}
Under the VSUR, $K_3$ reads
\begin{equation}
K_3=\frac{1}{16}+2 \cos (\Omega \tau)-\frac{5}{4} \cos (2 \Omega \tau)+\frac{3}{16} \cos (4 \Omega \tau) .
\end{equation}
We measure the magnitude of $K_3$ under different values of $\Omega\tau\in[0,2 \pi]$. 

In the trapped-ion system, we use the laser-ion interaction to control the ion qutrit. An ion with an energy level spacing $\omega_{0}$, interacting with a laser of the frequency $\omega_{l}$ forms a system, and the Hamiltonian raeds
\begin{eqnarray}
H_{L}=H_{0}+H_{I},
\end{eqnarray}
where $H_{0}$ describes the energy of the ion itself and $H_{I}$ describes the laser-ion interaction. The coupling strength is given by Rabi frequency $\Omega_{R}$. $H_{0}$ and $H_{I}$ can be expressed as
\begin{equation}
H_{0}=\frac{1}{2} \hbar \omega_{0} \sigma_{z},
\end{equation}
and
\begin{equation}
H_{I}=\frac{1}{2} \hbar \Omega_{R}\left(\sigma^{+}+\sigma^{-}\right)\left(e^{i\left(\omega_{l} t+\phi\right)}+e^{-i\left(\omega_{l} t+\phi\right)}\right).
\end{equation}
Here, $\phi$ denotes the initial phase of the laser, $t$ denotes the duration of interaction between the laser and the ion, $\sigma_{z}$ denotes the $z$ component of the Pauli operator, and $\sigma^{+}$ ($\sigma^{-}$) is the spin-flip operator. In the rotating frame with $\omega_{0}=\omega_{l}$, the interaction Hamiltonian can be written as
\begin{equation}
H_{I}=\frac{1}{2} \hbar \Omega_{R}\left(e^{i \phi} \sigma^{+}+e^{-i \phi} \sigma^{-}\right).
\end{equation}
The evolution operator under $H_{I}$ is the rotation operation in a two-dimensional Hilbert space
\begin{equation}
R(\theta, \phi)=\left(\begin{array}{cc}
\cos \frac{\theta}{2} & -i \sin \frac{\theta}{2} e^{-i \phi} \\
\hspace{0.6em}
-i \sin \frac{\theta}{2} e^{i \phi} & \cos \frac{\theta}{2}
\end{array}\right), \label{thisEq}
\end{equation}
where $\theta$ and $\phi$ denote the angle and phase of the rotation operation. All operations in the experiment must formally satisfy Eq.\,(\ref{thisEq}). A unitary matrix which acts on a high-dimensional Hilbert space can be decomposed into a product of two-level rotation matrices which act non-trivially only on two vector components \cite{nielsen2002quantum}. For example, two-dimensional rotations in the three-dimensional space can be represented as the following forms
\begin{equation}
\begin{aligned}
R_{1}\left(\theta_{1}, \phi_{1}\right)=\left(\begin{array}{ccc}
\cos \frac{\theta_{1}}{2} & -i \sin \frac{\theta_{1}}{2} e^{-i \phi_{1}} & 0 \\
-i \sin \frac{\theta_{1}}{2} e^{i \phi_{1}} & \cos \frac{\theta_{1}}{2} & 0 \\
0 & 0 & 1
\end{array}\right)  ,\label{Eq16}
\end{aligned}
\end{equation}
\begin{equation}
\begin{aligned}
R_{2}\left(\theta_{2}, \phi_{2}\right)=\left(\begin{array}{ccc}
\cos \frac{\theta_{2}}{2} & 0 & -i \sin \frac{\theta_{2}}{2} e^{-i \phi_{2}} \\
0 & 1 & 0 \\
-i \sin \frac{\theta_{2}}{2} e^{i \phi_{2}} & 0 & \cos \frac{\theta_{2}}{2}
\end{array}\right) ,\label{Eq17}
\end{aligned}
\end{equation}
and 
\begin{equation}
\begin{aligned}
R_{3}\left(\theta_{3}, \phi_{3}\right)=\left(\begin{array}{ccc}
1 & 0 & 0 \\
0&\cos \frac{\theta_{3}}{2} & -i \sin \frac{\theta_{3}}{2} e^{-i \phi_{3}}  \\
0&-i \sin \frac{\theta_{3}}{2} e^{i \phi_{3}} & \cos \frac{\theta_{3}}{2} 
\end{array}\right)  ,\label{Eq18}
\end{aligned}
\end{equation}
respectively. To facilitate the decomposition process of the three-dimensional unitary matrix, we introduce an auxiliary dimension in our experiment, defined as the fourth dimension in the matrix. The specific use of the auxiliary dimension can be referred to in Appendix A. 

If the system's evolution time satisfies $\Omega\left(t_{\beta}-t_{\alpha}\right)=\pi$, the original three-dimensional evolution operator,  via expanding to four-dimensional Hilbert space, can be decomposed into the product of three matrices as
\begin{widetext}
\begin{equation}
U_{10}=\left(\begin{array}{cccc}
0 & 0 & -1 & 0 \\
0 & -1 & 0 & 0 \\
-1 & 0 & 0 & 0 \\
0 & 0 & 0 & 1
\end{array}\right)=\left(\begin{array}{cccc}
1 & 0 & 0 & 0 \\
0 & 1 & 0 & 0 \\
0 & 0 & -1 & 0 \\
0 & 0 & 0 & -1
\end{array}\right)\left(\begin{array}{cccc}
0 & 0 & -1 & 0 \\
0 & 1 & 0 & 0 \\
1 & 0 & 0 & 0 \\
0 & 0 & 0 & 1
\end{array}\right)\left(\begin{array}{cccc}
1 & 0 & 0 & 0 \\
0 & -1 & 0 & 0 \\
0 & 0 & 1 & 0 \\
0 & 0 & 0 & -1
\end{array}\right).
\end{equation}
\end{widetext}
If the evolution time does not satisfy $\Omega\left(t_{\beta}-t_{\alpha}\right)=\pi$, the high-dimensional matrix needs to be decomposed according to the method given in Ref.\,\cite{nielsen2002quantum}. Then a simple transformation process is performed to decompose original matrix into seven matrices (see Appendix A for a detailed decomposition process). 

\section{Experimental setup and results analysis}
In this work, the LGI test is demonstrated using a single ${ }^{40} \mathrm{Ca}^{+}$ ion trapped in a blade linear Paul trap \cite{siverns2017ion}. The ion needs to be cooled firstly. The cooling schemes used in our experiment include Doppler cooling \cite{eschner2003laser}, electromagnetically induced transparency (EIT) cooling \cite{roos2000experimental,morigi2000ground,zhang2021estimation}, and sideband cooling \cite{eschner2003laser}. After the whole cooling process, the mean phonon number of the ion can be brought down below 0.1. In addition, we trigger the experimental sequence synchronized to the 50 Hz frequency to prevent noise from the power supply \cite{schmidt2003coherence}.

\begin{figure}[t]
\includegraphics[width=0.5\textwidth]{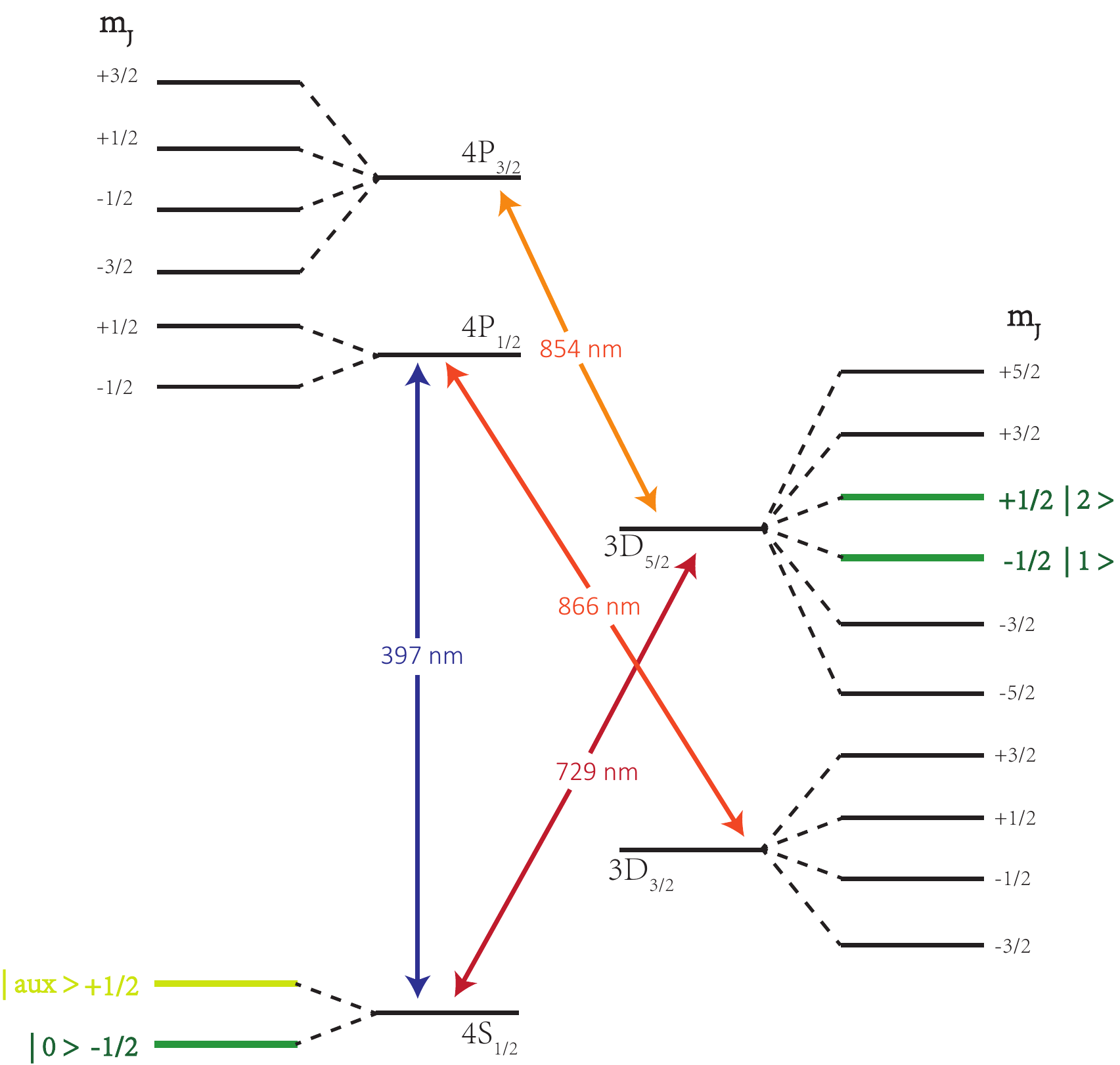}
\caption{\label{fig:2} Energy level diagram of ${ }^{40} \mathrm{Ca}^{+}$. The three levels $S_{1/2}(m_J=-1/2)$, $D_{5/2} (m_J=-1/2)$ and $D_{5/2}(m_J=+1/2)$ form a qutrit and $S_{1/2}(m_J=+1/2)$ is the auxiliary level we used.}
\end{figure}

The energy levels of ${ }^{40} \mathrm{Ca}^{+}$ have fine structure splitting under the magnetic field, as shown in Fig.\,\ref{fig:2}. We construct a qutrit using three levels of $\ket{0}=\ket{S_{1/2}(m_J=-1/2)}$, $\ket{1}=\ket{D_{5/2} (m_J=-1/2)}$ and $\ket{2}=\ket{D_{5/2}(m_J=+1/2)}$, and set the lowest energy level $\ket{0}$ as the initial state of our experiment. We select $\ket{aux}=\ket{S_{1/2}(m_J=+1/2)}$ as the auxiliary energy level used in the matrix decomposition. We define $Q=-1$ when the ion is in the state $\ket{0}$ and $Q=+1$ when the ion is in the state $\ket{1}$ or $\ket{2}$. 

The transition $S_{1/2}\leftrightarrow D_{5/2}$ involved in our experiment is controlled by a 729 nm light provided by the Titanium sapphire laser. The laser's frequency, phase, amplitude, and duration are controlled by an arbitrary waveform generator (AWG) via an acousto-optic modulator (AOM). We calibrate the transition frequencies and $\Omega_{R}$ of the four spectral lines $S_{1/2}(m_J=\pm 1/2)\leftrightarrow D_{5/2}(m_J=\pm 1/2)$ by scanning  laser frequency and duration. According to the specific $\tau $, we can determine the specific form of the evolution operators $U_{\beta \alpha}$ by Eq.\,(\ref{Eq8}). A set of $R(\theta ,\phi)$ can be obtained by matrix decomposition of the evolution operator. In the rotation operation $R(\theta ,\phi)$, $\theta $ is related to the laser duration, which can be determined by
\begin{equation}
t=\frac{\theta}{\Omega_{R}},
\end{equation}
and $\phi $ denotes the relative phase of the laser that can be set directly by the AWG. According to the Pound-Drever-Hall (PDH) error signal spectrum measured by a rf spectrum analyzer, the phase noise has been proved to be from the specific frequency band of Titanium sapphire laser \cite{zhang2021estimation}. To avoid the ac Stark frequency shift under the high power laser and the noise  mentioned above, we select $\Omega_{R}$ at about $2\pi \times$8 kHz in our experiment. 

\begin{figure}[b]
\includegraphics[width=0.37\textwidth]{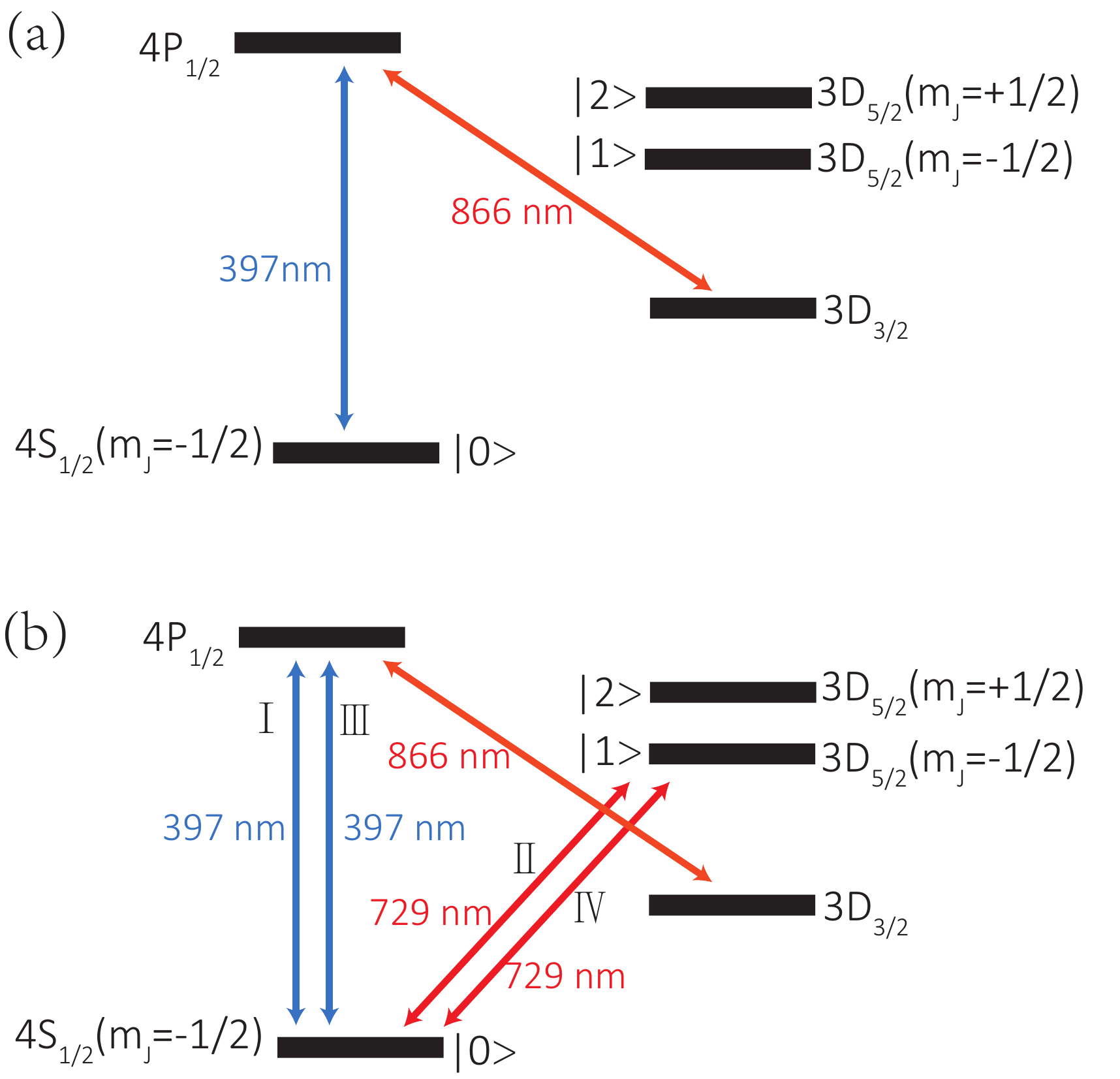}
\caption{\label{fig:1}  Schemes for two different measurement methods. (a) The measurement scheme using the LSUR. (b) The measurement scheme using the VSUR. The laser sequence used in the four-step operations of the measurement is shown by serial numbers \uppercase\expandafter{\romannumeral1}, \uppercase\expandafter{\romannumeral2}, \uppercase\expandafter{\romannumeral3}, and \uppercase\expandafter{\romannumeral4}.}
\end{figure}

The electron shelving technique \cite{nagourney1986shelved,sauter1986observation,bergquist1986observation} is used to discriminate between the state $\ket{0}$ and the state $\ket{1}$ or $\ket{2}$. The state $\ket{0}$ is detected when the photomultiplier (PMT) collects the fluorescence. No fluorescence can be observed if the ion is in the state $\ket{1}$ or $\ket{2}$ as the level $D_{5/2}$ is not coupled to the level $P_{1/2}$ by the 397 nm laser. Since the ion in the level $P_{1/2}$ has a probability of leaking to the level $D_{3/2}$, an 866 nm laser is needed to pump the ion back to the $S_{1/2}\leftrightarrow P_{1/2}$ transition cycle. In the LGI test under the LSUR, we define Q as 1 when the ion is in the state $\ket{1}$ or $\ket{2}$, so we only need to distinguish the state $\ket{0}$ and the state $\ket{1}$ or $\ket{2}$ by the electron shelving technique as shown in Fig.\,\ref{fig:1}(a). However, in the LGI test under the VSUR, we need to determine whether the ion is located in the state $\ket{0}$, the state $\ket{1}$, or the state $\ket{2}$. The sequence of four-step operations of the measurement is shown by serial numbers \uppercase\expandafter{\romannumeral1}, \uppercase\expandafter{\romannumeral2}, \uppercase\expandafter{\romannumeral3}, and \uppercase\expandafter{\romannumeral4} in Fig.\,\ref{fig:1}(b). We first distinguish whether the ion is in the state $\ket{0}$ by the electron shelving technique. Secondly, we exchange the state $\ket{0}$ and the state $\ket{1}$ with the 729 nm laser. Then we determine whether the ion is in the state $\ket{0}$ at this time. If so, the ion is in the state $\ket{1}$ before the state exchange; otherwise, it is in the state $\ket{2}$ before the state exchange. Finally, if the measurement is carried out before the second unitary evolution, we need to use the identical 729 nm laser to make the state of the ion return to the state before step \uppercase\expandafter{\romannumeral2} for the second unitary evolution by applying step \uppercase\expandafter{\romannumeral4}.

\begin{figure}[b]
\includegraphics[width=0.49\textwidth]{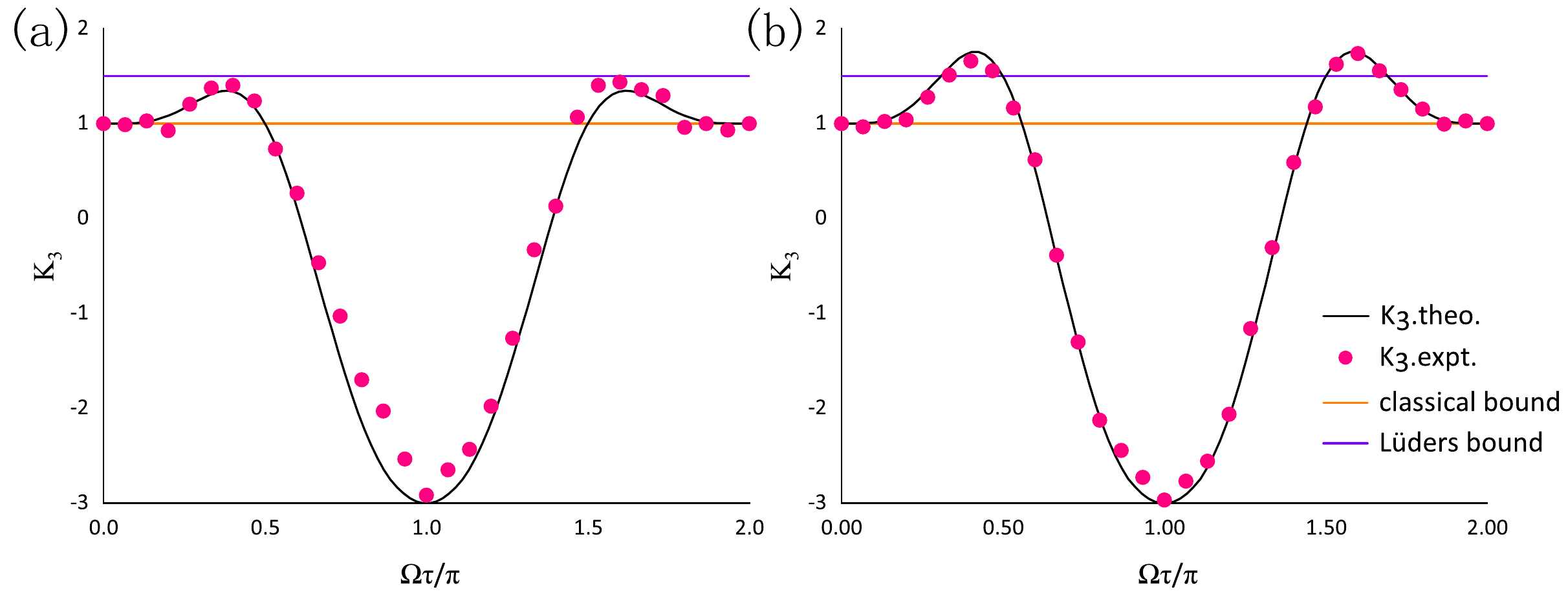}
\caption{\label{fig:5} Experimental results of $K_3$ under 31 different settings of $\Omega\tau\in[0,2 \pi]$. (a) The experimentally determined result under the LSUR. (b) The experimentally determined result under the VSUR. The black curve is the theoretical result, the red dots are the experimental results, the orange horizontal line is the classical bound, and the violet horizontal line is the Lüders bound. The error bars are not visible as the statistical error is about $10 ^{-2} $.}
\end{figure}

Each set of experiments is repeated at least ten thousand times to obtain the populations of ion state. According to the populations, we can calculate $C_{\beta\alpha}$, and then get $K_3$ by Eq.\,(\ref{con:K3}). The values of $K_3$ corresponding to different $\Omega\tau $ are shown in Fig.\,\ref{fig:5}. Fig.\,4(a) and  Fig.\,4(b) are the experimental results under the LSUR and VSUR respectively. The black curve is the theoretical result, the red dots are experimental results, the orange horizontal line is the CB, and the violet horizontal line is the LB. It can be seen that $K_3$ exceeds the CB regardless of which state update rule is used. It is not surprising that the LGI is violated as the state of a single ion definitely can be superposition. Under the same experimental settings, we discover that the $K_3$ does not exceed the LB under the LSUR, but it exceeds the LB under the VSUR. Under the VSUR, the theoretical maximum value is 1.7565 appearing at about $\Omega\tau = 1.585\pi$. The maximum value in our experiment is $1.739 \pm 0.014$ appearing at $\Omega\tau = 1.6\pi$, which exceeds the LB with a $17\sigma$ confidence level, and agrees with the theoretical value at $\Omega\tau = 1.6\pi$, which is 1.750. The difference of LGI's upper bound is related to the projection measurement protocol of two state update rules. The projection measurement protocol under the VSUR measures each one-dimensional space, which makes the wave function collapse completely, but the projection measurement protocol under the LSUR retains the coherence between the states $\ket{1}$ and $\ket{2}$. When the eigenvalue $Q =-1$, there is no difference between the projection measurement protocol under the LSUR and VSUR because there is no degeneracy. However, the situation is different when the eigenvalue $Q =+1$. $\Pi_{1}$ and $\Pi_{2}$ project the ion state to the states $\ket{1}$ and $\ket{2}$ respectively. According to the LSUR, after one measurement, the change of the density operator can be described as
\begin{equation}
\begin{aligned}
\rho_L&=\Pi_{+} \rho_0 \Pi_{+}=\left(\Pi_1+\Pi_2\right) \rho_0\left(\Pi_1+\Pi_2\right)\\
&=\Pi_1 \rho_0 \Pi_1+\Pi_2 \rho_0 \Pi_2+\Pi_1 \rho_0 \Pi_2+\Pi_2 \rho_0 \Pi_1 .
\end{aligned}
\end{equation}
However, the change of the density operator under the VSUR is described as
\begin{equation}
\begin{aligned}
\rho_V&=\Pi_1 \rho_0 \Pi_1+\Pi_2 \rho_0 \Pi_2 \\
&=\rho_L-\left(\Pi_1 \rho_0 \Pi_2+\Pi_2 \rho_0 \Pi_1\right).
\end{aligned}
\end{equation}
The term $-\left(\Pi_1 \rho_0 \Pi_2+\Pi_2 \rho_0 \Pi_1\right)$ is responsible for the violation of the LB.

The experimental error in our LGI test mainly comes from the following four factors. Firstly, the decoherence of the ion contributes to the deviation between theoretical and experimental results. It is worth mentioning that the measurement under the VSUR has more experimental steps but the experimental results under the VSUR are more consistent with the theory. Unlike the measurement protocol under the VSUR, the measurement under the LSUR retains the coherence between the states $\ket{1}$ and $\ket{2}$, so the coherence time effects more on the measurement under the LSUR. Secondly, according to data from the experimental calibration process, the fidelity of the evolution operations is slightly worse at low power of the 729 nm laser. There will be a large ac Stark frequency shift when the 729 nm laser power is high; on the other hand, the phase noise will increase with the decrease of the laser power due to the locking circuit of this narrow line-width laser. The ac Stark frequency shift will affect the energy level spacing in our experiment, and the phase noise will affect the purity of the 729 nm laser. To trade off the influence of the ac Stark frequency shift and the phase noise, we choose the appropriate laser power. $\Omega_{R}$ of four spectral lines in our experiment are about $2\pi \times$8kHz. With this $\Omega_{R}$, the influence of the ac Stark frequency shift can be almost ignored, and the evolution fidelity is about $98\%$. Thirdly, the fidelity of our initial state preparation is about $99.4\%$ due to the polarization of the laser is not pure, and it is difficult to make the laser parallel absolutely with the magnetic field. Fourthly, during this period of our experiment, the power of the 397 nm laser slightly fluctuates. Since the electron shelving technique is implemented by a 397 nm laser, the fluctuation may slightly affect the measurement fidelity, and the 397 nm laser instability affects the effect of Doppler cooling too.

\section{Conclusion}
We have realized the LGI test in a natural three-level trapped-ion system, and obtained $K_3$ under the LSUR and VSUR respectively, which are consistent with the theoretical predictions. We directly compared the upper bound of LGI under two state update rules in the same experimental system for the first time. We adopted the model of a large spin precession in the magnetic field, and experimentally obtained by far the most significant violation under the VSUR, which benefits from the high-fidelity operations and long coherence time of the trapped-ion system. Our experiment can be scaled to the experiment with more dimensions or more measurement moments. On the one hand, more operational energy levels can be introduced based on the Zeeman splitting structure of ${ }^{40} \mathrm{Ca}^{+}$. There are five energy levels in the  $D_{5/2}$ that can be coupled with the level $S_{1/2}(m_J=-1/2)$. Therefore, our experiment can be extended up to a six-dimensional system. On the other hand, We can add more unitary evolutions and measurements, so our experiment can also be extended to the experiment with more measurement moments.

\begin{acknowledgments}
This work is supported by the National Natural Science Foundation of China under Grants No.\,11904402, No.\,12004430, No.\,12074433, No.\,12174447, No.\,12174448 and No.\,12204543.
\end{acknowledgments}

\appendix

\section{the matrix decomposition}

The corresponding relationship between the energy levels and the elements $M_{qp}$ of the four-dimensional matrix is shown in Tab.\,\ref{tab:table3}. Three experimental levels $S_{1/2}(m_J=-1/2)$, $D_{5/2}(m_J=-1/2)$, $D_{5/2}(m_J=+1/2)$, and an auxiliary level $S_{1/2}(m_J=+1/2)$ correspond to the first, second, third and fourth dimension of the matrix, respectively.

\begin{table*}
\caption{\label{tab:table3}The corresponding relationship between the energy levels and matrix elements $M_{qp}$.}
\begin{ruledtabular}
\begin{tabular}{ccccc}
 &$S_{1/2}(m_J=-1/2)$&$D_{5/2}(m_J=-1/2)$&$D_{5/2}(m_J=+1/2)$
&$S_{1/2}(m_J=+1/2)$\\ \hline
 $S_{1/2}(m_J=-1/2)$&$M_{11}$&$M_{12}$&$M_{13}$&$M_{14}$ \\
 $D_{5/2}(m_J=-1/2)$&$M_{21}$&$M_{22}$&$M_{23}$&$M_{24}$\\
 $D_{5/2}(m_J=+1/2)$&$M_{31}$&$M_{32}$&$M_{33}$&$M_{34}$\\
 $S_{1/2}(m_J=+1/2)$&$M_{41}$&$M_{42}$&$M_{43}$&$M_{44}$\\
\end{tabular}
\end{ruledtabular}
\end{table*}

The $x$ component of the spin angular momentum in the three-dimensional space is
\begin{equation}
J_x=\frac{1}{\sqrt{2}} \left(\begin{array}{ccc}
0 & 1 & 0 \\
1 & 0 & 1 \\
0 & 1 & 0
\end{array}\right).
\end{equation}

When $\Omega\left(t_{\beta}-t_{\alpha}\right) \neq \pi$, let $\Omega\tau=\epsilon$, then the unitary evolution operator can be written as
\begin{equation}
\begin{split}
U_{\beta \alpha}
& =e^{-i H\left(t_{\beta}-t_{\alpha}\right)}=e^{-i J_{\mathrm{x}} \Omega\left(t_{\beta}-t_{\alpha}\right)}=e^{-i J_{\mathrm{x}} \epsilon}\\
& =\left(\begin{array}{ccc}
\frac{1}{2}+\frac{\cos \epsilon}{2} & \frac{-i \sin \epsilon}{\sqrt{2}} & -\frac{1}{2}+\frac{\cos \epsilon}{2} \\
\frac{-i \sin \epsilon}{\sqrt{2}} & \cos \epsilon & \frac{-i \sin \epsilon}{\sqrt{2}} \\
-\frac{1}{2}+\frac{\cos \epsilon}{2} & \frac{-i \sin \epsilon}{\sqrt{2}} & \frac{1}{2}+\frac{\cos \epsilon}{2}
\end{array}\right).
\end{split}
\end{equation}\\
According to the method in Ref.\,\cite{nielsen2002quantum}, $U_{\beta \alpha}$ can be decomposed into
\begin{equation}
\begin{split}
U_{\beta \alpha}
& =\left(\begin{array}{ccc}
\frac{\sqrt{2}(1+\cos \epsilon)}{\sqrt{5+4 \cos \epsilon-\cos 2 \epsilon}} & \frac{2 i \sin \epsilon}{\sqrt{5+4 \cos \epsilon-\cos 2 \epsilon}} & 0 \\
-\frac{2 i \sin \epsilon}{\sqrt{5+4 \cos \epsilon-\cos 2 \epsilon}} & -\frac{\sqrt{2}(1+\cos \epsilon)}{\sqrt{5+4 \cos \epsilon-\cos 2 \epsilon}} & 0 \\
0 & 0 & 1
\end{array}\right) \\ 
& \left(\begin{array}{ccc}
\frac{\cos ^{2} \frac{\epsilon}{2} \sqrt{3-\cos \epsilon}}{\sqrt{2}\left|\cos \frac{\epsilon}{2}\right|} & 0 & \frac{1}{2}(-1+\cos \epsilon) \\
0 & 1 & 0 \\
\frac{1}{2}(-1+\cos \epsilon) & 0 & -\frac{\cos ^{2} \frac{\epsilon}{2} \sqrt{3-\cos \epsilon}}{\sqrt{2}\left|\cos \frac{\epsilon}{2}\right|}
\end{array}\right) \\
& \left(\begin{array}{ccc}
1 & 0 & 0 \\
0 & \frac{-1-\cos \epsilon}{\left|\cos \frac{\epsilon}{2}\right| \sqrt{6-2 \cos \epsilon}} & \frac{2 i \sin \epsilon}{\sqrt{5+4 \cos \epsilon-\cos 2 \epsilon}} \\
0 & \frac{2 i \sin \epsilon}{\sqrt{5+4 \cos \epsilon-\cos 2 \epsilon}} & \frac{-1-\cos \epsilon}{\left|\cos \frac{\epsilon}{2}\right| \sqrt{6-2 \cos \epsilon}}
\end{array}\right).
\end{split}
\end{equation}
We note that these three matrices can not be directly implemented in experiments, the first two matrices still have differences in positive and negative numbers with \begin{equation}
R(\theta, \varphi)=\left(\begin{array}{cc}
\cos \frac{\theta}{2} & -i \sin \frac{\theta}{2} e^{-i \varphi} \\
\hspace{0.6em}
-i \sin \frac{\theta}{2} e^{i \varphi} & \cos \frac{\theta}{2}
\end{array}\right), \label{con:inventoryflow}
\end{equation}
and the third matrix involves an infeasible coupling between $D_{5/2}(m_J=-1/2)$ and $D_{5/2}(m_J=+1/2)$. We extend it to four dimensions and make simple transformations
\begin{widetext}
\begin{equation}
\begin{split}
U_{\beta \alpha} 
& =\left(\begin{array}{cccc}
\frac{\sqrt{2}(1+\cos \epsilon)}{\sqrt{5+4 \cos \epsilon-\cos 2 \epsilon}} & \frac{2 i \sin \epsilon}{\sqrt{5+4 \cos \epsilon-\cos 2 \epsilon}} & 0 & 0 \\
-\frac{2 i \sin \epsilon}{\sqrt{5+4 \cos \epsilon-\cos 2 \epsilon}} & -\frac{\sqrt{2}(1+\cos \epsilon)}{\sqrt{5+4 \cos \epsilon-\cos 2 \epsilon}} & 0 & 0 \\
0 & 0 & 1 & 0 \\
0 & 0 & 0 & 1
\end{array}\right) \\ 
& \left(\begin{array}{cccc}
\frac{\cos ^{2} \frac{\epsilon}{2} \sqrt{3-\cos \epsilon}}{\sqrt{2}\left|\cos \frac{\epsilon}{2}\right|} & 0 & \frac{1}{2}(-1+\cos \epsilon) & 0 \\
0 & 1 & 0 & 0 \\
\frac{1}{2}(-1+\cos \epsilon) & 0 & -\frac{\cos ^{2} \frac{\epsilon}{2} \sqrt{3-\cos \epsilon}}{\sqrt{2}\left|\cos \frac{\epsilon}{2}\right|} & 0 \\
0 & 0 & 0 & 1
\end{array}\right)\left(\begin{array}{cccc}
1 & 0 & 0 & 0 \\
0 & \frac{-1-\cos \epsilon}{\left|\cos \frac{\epsilon}{2}\right| \sqrt{6-2 \cos \epsilon}} & \frac{2 i \sin \epsilon}{\sqrt{5+4 \cos \epsilon-\cos 2 \epsilon}} & 0 \\
0 & \frac{2 i \sin \epsilon}{\sqrt{5+4 \cos \epsilon-\cos 2 \epsilon}} & \frac{-1-\cos \epsilon}{\left|\cos \frac{\epsilon}{2}\right| \sqrt{6-2 \cos \epsilon}} & 0 \\
0 & 0 & 0 & 1
\end{array}\right) \\
& =\left(\begin{array}{cccc}
1 & 0 & 0 & 0 \\
0 & -1 & 0 & 0 \\
0 & 0 & 1 & 0 \\
0 & 0 & 0 & -1
\end{array}\right)\left(\begin{array}{cccc}
\frac{\sqrt{2}(1+\cos \epsilon)}{\sqrt{5+4 \cos \epsilon-\cos 2 \epsilon}} & \frac{2 i \sin \epsilon}{\sqrt{5+4 \cos \epsilon-\cos 2 \epsilon}} & 0 & 0 \\
\frac{2 i \sin \epsilon}{\sqrt{5+4 \cos \epsilon-\cos 2 \epsilon}} & \frac{\sqrt{2}(1+\cos \epsilon)}{\sqrt{5+4 \cos \epsilon-\cos 2 \epsilon}} & 0 & 0 \\
0 & 0 & 1 & 0 \\
0 & 0 & 0 & 1
\end{array}\right)\left(\begin{array}{cccc}
1 & 0 & 0 & 0 \\
0 & 1 & 0 & 0 \\
0 & 0 & -1 & 0 \\
0 & 0 & 0 & -1
\end{array}\right) \\
& \left(\begin{array}{cccc}
\frac{\cos ^{2} \frac{\epsilon}{2} \sqrt{3-\cos \epsilon}}{\sqrt{2}\left|\cos \frac{\epsilon}{2}\right|} & 0 & \frac{1}{2}(-1+\cos \epsilon) & 0 \\
0 & 1 & 0 & 0 \\
-\frac{1}{2}(-1+\cos \epsilon) & 0 & \frac{\cos ^{2} \frac{\epsilon}{2} \sqrt{3-\cos \epsilon}}{\sqrt{2}\left|\cos \frac{\epsilon}{2}\right|} & 0 \\
0 & 0 & 0 & 1
\end{array}\right)\left(\begin{array}{cccc}
0 & -1 & 0 & 0 \\
1 & 0 & 0 & 0 \\
0 & 0 & 1 & 0 \\
0 & 0 & 0 & 1
\end{array}\right)\left(\begin{array}{cccc}
\frac{-1-\cos \epsilon}{\left|\cos \frac{\epsilon}{2}\right| \sqrt{6-2 \cos \epsilon}} & 0 & \frac{2 i \sin \epsilon}{\sqrt{5+4 \cos \epsilon-\cos 2 \epsilon}} & 0 \\
0 & 1 & 0 & 0 \\
\frac{2 i \sin \epsilon}{\sqrt{5+4 \cos \epsilon-\cos 2 \epsilon}} & 0 & \frac{-1-\cos \epsilon}{\left|\cos \frac{\epsilon}{2}\right| \sqrt{6-2 \cos \epsilon}} & 0 \\
0 & 0 & 0 & 1
\end{array}\right)\left(\begin{array}{cccc}
0 & 1 & 0 & 0 \\
-1 & 0 & 0 & 0 \\
0 & 0 & 1 & 0 \\
0 & 0 & 0 & 1
\end{array}\right).
\end{split}
\end{equation}
\end{widetext}

\nocite{*}

\bibliographystyle{unsrt}
\bibliography{apssamp}

\end{document}